\documentclass[prb, twocolumn, a4paper, superscriptaddress, showpacs, amsmath, amssymb,
nofootinbib, floatfix] {revtex4}

\usepackage{graphicx}
\usepackage{bm}

\begin{document}

\title{Evolution of the correlation functions in 2D dislocation systems}

\author{P\'eter Dus\'an Isp\'anovity}
\email{ispanovity@metal.elte.hu}
\affiliation{Department of Materials Physics, E\"otv\"os University,
H-1117 Budapest, P\'azm\'any s\'et\'any 1/a, Hungary}

\author{Istv\'an Groma}
\email{groma@metal.elte.hu}
\affiliation{Department of Materials Physics, E\"otv\"os University,
H-1117 Budapest, P\'azm\'any s\'et\'any 1/a, Hungary}

\author{G\'eza Gy\"orgyi}
\email{gyorgyi@glu.elte.hu}
\affiliation{Institute of Theoretical Physics, HAS Research Group, E\"otv\"os University,
H-1117 Budapest, P\'azm\'any s\'et\'any 1/a, Hungary}
\affiliation{Research Institute for Solid State Physics and Optics, H-1525
Budapest, P.O.\ Box 49, Hungary}

\date{\today}

\begin{abstract}

In this paper spatial correlations of parallel edge dislocations are studied. After closing a hierarchy of equations for the many-particle density functions by the Kirkwood superposition approximation, we derive evolution equations for the correlation functions. It is found that these resulting equations and those governing the evolution of density fields of total as well as geometrically necessary dislocations around a single edge dislocation are formally the same. The second case corresponds to the already described phenomenon of Debye screening of an individual dislocation. This equivalence of the correlation functions and screened densities is demonstrated also by discrete dislocation dynamics simulation results, which confirm the physical correctness of the applied Kirkwood superposition approximation. Relation of this finding and the linear response theory in thermal systems is also discussed.

\end{abstract}


\maketitle

\section{Introduction}

It is well known that plastic deformation of crystalline materials is caused by the motion of a large number of interacting dislocations. Although the properties of individual dislocations have been known for decades, the question how to build up a micron-scale level continuum description of dislocations is still an unresolved issue in the theory of crystal plasticity. The appropriate continuum theory could provide a framework for understanding dislocation pattern formation or size-effects.

Several phenomenological models have been proposed for dislocation patterning so far. Here we first mention the work of Walgraef and Aifantis\cite{walgraef1, walgraef2}, who adopted a reaction-diffusion model from the description of oscillating chemical reactions. In their model second-order gradient terms are introduced to account for the spatial fluctuations of the dislocation density. Although the framework turned out to be very successful in the prediction of different dislocation structures, the physical origin of the length scales introduced through the coefficients of the gradient terms is not really understood. The same problem applies to the model of Kratochv\'il and co-workers, in which nonlocal terms are introduced for describing spatial interactions through the sweeping mechanism.\cite{kratochvil1, kratochvil2}

In a non-phenomenological model proposed by Groma a system of parallel edge dislocations in single slip was studied.\cite{groma1} By performing ensemble averaging on statistically equivalent dislocation distributions a chain of equations for the many-body dislocation density fields was derived. First, dislocation correlations were neglected, which corresponds to a mean-field approximation.\cite{groma4} Here the stress acting on dislocations is simply the ``self-consistent'' field, which is the long-range stress field of the geometrically necessary dislocations. If the dislocation system is correlated, which is indeed the case for real systems (see, e.g.,\ Ref.\ \onlinecite{zaiser1}), new stress-like terms appear in the constitutive equations.\cite{groma2} The most important non-trivial term is the gradient-like ``back-stress'', which can be interpreted as the short-range effect of dislocation pile-ups. In order to check the validity of this two-dimensional (2D) theory, its predictions were compared with discrete dislocation dynamics simulation results and good agreement was found.\cite{yefimov1, yefimov2}

There are several current approaches in the literature to develop a density-based continuum model for three dimensions (3D). El-Azab based his theory on Nye's dislocation density tensor.\cite{elazab1} After studying the mean-field approximation of an uncorrelated ensemble, in their recent paper El-Azab \emph{et al}.\ turned to the investigation of dislocation correlations on simulated 3D dislocation ensembles in a body-centered-cubic crystal.\cite{elazab2, elazab3} Csikor \textit{et al}.\ reported about a complementary numerical study focused on the range of 3D dislocation pair correlations in face-centered-cubic materials.\cite{csikor1} Recently, another promising theory has been proposed by Hochrainer and co-workers who studied the evolution of dislocation lines in a higher-dimensional space.\cite{zaiser2, hochrainer, zaiser3} Motivated by the 2D results dislocation correlations are introduced into the theory semi-phenomenologically by considering short-range stress terms such as the aforementioned back-stress.\cite{zaiser3} Similar extension was adopted by Schwarz \emph{et al}.\ too\cite{schwarz} for a continuum theory of interacting curved dislocations\cite{sedlacek}. In all mentioned current 3D continuum theories of dislocation dynamics one of the biggest challenges is the precise incorporation of the effects of dislocation correlations, which seems to be unavoidable. So, the investigation of the evolution of the correlation properties is of great importance, but even in 2D this problem has not been solved yet.

In the past two years new light has been shed on equilibrated dislocation systems by the discovery of the effect of dislocation screening. \cite{groma3,groma5} Its most direct physical significance is that it explains the extensivity of the elastic energy in equilibrium by the appropriate relative arrangement of the dislocations, resulting generically in a finite interaction length rather than the unscreened infinite range interaction. On the technical side, in the simplest example of single slip, 2D, edge dislocation systems, an effective thermodynamic potential was proposed for the purpose of the variational calculation of the geometrically necessary density. Summarizing the results, for large and constant total dislocation density and small but variable geometrically necessary density, the equation for the induced density with arbitrary external dislocations was constructed and its Green function for infinite boundary conditions was given analytically. Hence the induced density and the resulting screened elastic potential by a single external dislocation was obtained, showing suppression in all directions exponentially strong except for one axis, where it was of power type. This phenomenon bears close resemblance to Debye screening of an external charge in Coulomb systems, so this term was adopted also for dislocations. In the same work\cite{groma3} numerical results were presented, but because of the prohibitive fluctuations of the dislocation distribution induced by an external, fixed dislocation, rather the signed pair correlation of dislocations in equilibrium was monitored. Even in the latter function an agreement with the theory was found.

It should be borne in mind that the relation between the externally induced density (response function) and the correlation function is based on linear response theory in thermal systems. However, in the dislocation system considered presently entropic effects are suppressed, i.e., thermal fluctuations are negligible beside the Peach-Koehler forces. Therefore, traditional arguments of linear response theory do not obviously apply. This raises the question why then the externally induced dislocations obey properties similar to those of correlation functions. The problem can be actually posed in a broad sense, namely, to what extent the response function to an external particle and the two-particle correlation function, in or off equilibrium, are similar. The evolution equations for the one-point densities, with some simplifying hypotheses, are known, \cite{groma2} and they have been linked to the variational approach by a phase field construction\cite{groma5}. The evolution equations for the two-point correlation functions, however, could not be brought to any treatable form so far.

As a recent development, Vinogradov and Willis published an evolution equation for the correlation function of long parallel screw dislocations with the same Burgers vectors.\cite{vinogradov1,vinogradov2} They even solved the resulting equation for the static case analytically. It was found, however, that the generalization of this result for the case of two possible Burgers vectors (positive and negative dislocations), or for edge dislocations, leads to a solution not decaying at infinity, thus cannot be correct physically.

In this paper we investigate the evolution of correlation functions of long parallel edge dislocations with single slip. We start our analysis with the Bogoliubov-Born-Green-Kirkwood-Yvon (BBGKY) hierarchy, describing the evolution of dislocation many-body densities, which was derived earlier by Groma.\cite{groma1} In contrast to the cluster approximation used by Vinogradov and Willis,\cite{vinogradov1,vinogradov2} we apply the Kirkwood superposition approximation as a method to close the chain of equations. The resulting equations, when rewritten for appropriately introduced single-point fields, are similar to those describing the evolution of one-dislocation density functions in the presence of a single external dislocation. In fact they become identical, if the external perturbation is small enough, thus we arrive at the analog of linear response theory in thermal systems, now obtained for density fields at zero temperature. It should be kept in mind, however, that the relation is now conditioned upon the Kirkwood approximation for the two-point correlations. In order to test the extent of the analogy between the one- and two-point densities, we also performed extensive simulations. First we compared the equilibrium distributions induced by the screened single external dislocation with the appropriate fields from the two-point correlations, and found a quite satisfactory agreement. Second, the time evolutions of the two types of density fields were matched and again close similarity was observed even before reaching equilibrium. This essentially demonstrates that the Kirkwood closure approximation was justified and the resulting equations indeed well describe the evolution of the correlation functions. We also analyze the cluster approximation of Vinogradov and Willis, adopted for edge dislocations, and show that it leads to a physically unacceptable result for the evolution equation of the correlation functions.

The paper is organized as follows. In Sec.\ \ref{sec:time_evol} we derive the new equations of motion for the two-point correlation functions and for the formally introduced, effectively one-point, fields. Section \ref{sec:debye} recuperates the equations for the single dislocation densities, while Sec.\ \ref{sec:conn} is devoted to the comparison of the previous two sections' results. Section \ref{sec:num_res} contains the presentation of the outcome of the simulations. The Conclusions are followed by the Appendix, containing background calculations for Sec.\ \ref{sec:time_evol}.

\section{Time evolution of the correlation functions}
\label{sec:time_evol}

Let us consider a system of parallel edge dislocations with single slip system, where only overdamped glide motion is allowed. By considering a cross-section of the crystal perpendicular to the dislocations the problem becomes two dimensional with point-like objects. If $s_i$ denotes the sign of the $i$th dislocation, then its Burgers vector is $\bm b_i = s_i \bm b$ [$s_i \in \{+, -\}$ and $\bm b := (b, 0)$].

In order to describe a discrete system of dislocations, it is useful to introduce discrete $n$-particle dislocation densities, e.g., $\rho_1^{D,s}(\bm r, t) := \sum_{i=1}^N \delta_{s, s_i} \delta(\bm r - \bm r_i(t))$ for $n=1$ in a system of $N$ dislocations (for details see, e.g., Ref. \onlinecite{groma5}). Here $\bm r_i(t)$ denotes the position of the $i$th dislocation, and with the superscript $s$ we make a distinction between the densities of positive and negative dislocations. Higher order discrete densities could be defined accordingly. In order to get smooth density functions, one has to perform averaging over statistically equivalent systems, which is denoted as, e.g., $\rho_{1}^s(\bm r, t) := \langle \rho_1^{D,s}(\bm r, t) \rangle$. After performing the averaging procedure, Groma obtained a chain of evolution equations for the $\rho_n^{s_1, \dots,s_n}$ $n$-particle density functions of the system.\cite{groma1} These equations give the rate change of $\rho_n^{s_1, \dots,s_n}$, which depends on both itself and the one-level higher density function $\rho_{n+1}^{s_1, \dots,s_{n+1}}$. This type of chain of equations is called BBGKY hierarchy in statistical physics. To make it possible to solve these infinite number of equations the hierarchy must be chopped at a certain level.

We start our investigations at the first level of the hierarchy, which governs the evolution of the one-particle density functions:\cite{groma1}
\begin{eqnarray}
&&\frac{\partial \rho_1^{s_1}(\bm r_1, t)}{\partial t} \nonumber \\
&& \ + \frac{\partial}{\partial \bm r_1} \sum_{s_2=\pm 1} \int\limits_{\mathbb R^2} \rho_2^{s_1,s_2}(\bm r_1, \bm r_2, t) \bm F^{s_1,s_2}(\bm r_1 - \bm r_2) d^2 r_2 \nonumber \\
&& \ + \frac{\partial}{\partial \bm r_1} \left[ \rho_1^{s_1}(\bm r_1, t) \bm F_\text{ext}^{s_1}(\bm r_1) \right] = 0, \label{eqn:one_part_evol}
\end{eqnarray}
where $\rho_n^{s_1, \dots ,s_n}$ denotes the $n$-particle dislocation density function with $s_i$ referring to the sign of the $i$th dislocation, $\bm F^{s_1,s_2}(\bm r_1 - \bm r_2)$ is the interaction force between two dislocations at positions $\bm r_1$ and $\bm r_2$, and with signs $s_1$ and $s_2$ ($\bm r_i \in \mathbb{R}^2$):
\begin{equation}
\bm F^{s_1, s_2}(\bm r) = s_1 s_2 \tau_\text{ind}(\bm r) \bm b.
\end{equation}
Here $\tau_\text{ind}$ is the shear stress field of a single edge dislocation positioned in the origin:
\begin{equation}
\tau_\text{ind}(\bm r) = \frac{\mu | \bm b |}{2 \pi (1-\nu)} \frac{x(x^2-y^2)}{(x^2+y^2)^2}
\end{equation}
in which $\mu$ and $\nu$ denote the shear modulus and the Poisson's ratio, respectively. A $\tau_\text{ext}(\bm r)$ spatially varying external shear stress applied to the system results in the extra $\bm F_\text{ext}^{s}(\bm r)$ force, which depends on the sign $s$ of the dislocation:
\begin{equation}
\bm F_\text{ext}^{s}(\bm r) = s \tau_\text{ext}(\bm r) \bm b.
\end{equation}
Solving (\ref{eqn:one_part_evol}) generally (with arbitrary boundary conditions) requires the additional knowledge of $\rho_2$.

In this section we investigate the relaxation of dislocations from a random initial state in an infinite medium at zero external shear stress. Since the equations of motion of the dislocations don't have a direct spatial dependence, the system is invariant under translations. In this case the system is homogeneous, meaning the one-particle density functions $\rho_1^s$ cannot have a spatial dependence, i.e.,
\begin{equation}
\rho^\pm := \rho_1^\pm(\bm r, t) = \text{const}.
\label{eqn:const_rho}
\end{equation}
This is a trivial solution of Eq.\ (\ref{eqn:one_part_evol}). One can draw two simple conclusions from it:
\begin{itemize}
\item The two-particle density functions depend only on the difference of their arguments:
\begin{equation}
\rho_2^{s_1,s_2}(\bm r_1, \bm r_2, t) = \rho_2^{s_1,s_2}(\bm r_1 - \bm r_2, t), \quad s_1,s_2 \in \{+,-\},
\label{eqn:rho_2_rel_coords}
\end{equation}
\item and the following identity holds:
\begin{equation}
\sum_{s_2=\pm 1} s_2 \int\limits_{\mathbb R^2} \rho_2^{s_1,s_2}(\bm r, t) \tau_\text{ind}(\bm r) d^2 r = 0, \quad s_1 \in \{+,-\}.
\end{equation}
\end{itemize}

We now proceed to the evolution equations of the two-particle density functions:\cite{groma1}
\begin{eqnarray}
&& \frac{\partial \rho_2^{s_1,s_2}(\bm r_1, \bm r_2, t)}{\partial t} \nonumber \\
&& \enskip + \left( \frac{\partial}{\partial \bm r_1} - \frac{\partial}{\partial \bm r_2} \right) \left[ \rho_2^{s_1, s_2}(\bm r_1, \bm r_2, t) \bm F^{s_1, s_2}(\bm r_1 - \bm r_2) \right] \nonumber \\
&& \enskip + \frac{\partial}{\partial \bm r_1} \sum_{s_3=\pm 1} \int\limits_{\mathbb R^2} \rho_3^{s_1, s_2, s_3}(\bm r_1, \bm r_2, \bm r_3, t) \bm F^{s_1, s_3}(\bm r_1 - \bm r_3) d^2 r_3 \nonumber \\
&& \enskip + \frac{\partial}{\partial \bm r_2} \sum_{s_3=\pm 1} \int\limits_{\mathbb R^2} \rho_3^{s_1, s_2, s_3}(\bm r_1, \bm r_2, \bm r_3, t) \bm F^{s_2, s_3}(\bm r_2 - \bm r_3) d^2 r_3 \nonumber \\
&& \enskip + \frac{\partial}{\partial \bm r_1} \left[ \rho_2^{s_1,s_2}(\bm r_1, \bm r_2, t) \bm F_\text{ext}^{s_1}(\bm r_1) \right] \nonumber \\
&& \enskip + \frac{\partial}{\partial \bm r_2} \left[ \rho_2^{s_1,s_2}(\bm r_1, \bm r_2, t) \bm F_\text{ext}^{s_2}(\bm r_2) \right] = 0. \label{eqn:two_part_evol}
\end{eqnarray}
It has to be noted that by definition $\rho_2^{s_1,s_2}(\bm r_1, \bm r_2, t) = \rho_2^{s_2,s_1}(\bm r_2, \bm r_1, t)$.

A possible method to truncate the hierarchy at this level is to express $\rho_3^{s_1,s_2,s_3}$ in the so-called cluster expansion, and neglect the three-particle correlation term, as it was done by Groma\cite{groma1}, and more recently by Vinogradov and Willis\cite{vinogradov1,vinogradov2}. This means $\rho_3^{s_1, s_2, s_3}$ is approximated as
\begin{eqnarray}
&& \rho_3^{s_1, s_2, s_3}(\bm r_1, \bm r_2, \bm r_3, t) \nonumber \\
&& \quad \approx \rho_1^{s_1}(\bm r_1, t) \rho_1^{s_2}(\bm r_2, t) \rho_1^{s_3}(\bm r_3, t) \nonumber \\
&& \quad \quad \times [ 1 + d^{s_1,s_2}(\bm r_1, \bm r_2, t) + d^{s_2,s_3}(\bm r_2, \bm r_3, t) \nonumber \\
&& \quad \quad \quad + d^{s_3,s_1}(\bm r_3, \bm r_1, t) ], \label{eqn:cluster}
\end{eqnarray}
where the dislocation-dislocation correlation functions are defined in the usual way:
\begin{equation}
d^{s_1,s_2}(\bm r_1, \bm r_2, t) := \frac{\rho_2^{s_1,s_2}(\bm r_1, \bm r_2, t)}{\rho_1^{s_1}(\bm r_1, t) \rho_1^{s_2}(\bm r_2, t)} - 1.
\label{eqn:corr}
\end{equation}
Substituting the cluster expansion (\ref{eqn:cluster}) into the second-order evolution equation (\ref{eqn:two_part_evol}) leads to a closed evolution equation for $\rho_2^{s_1,s_2}$, or equivalently, for $d^{s_1,s_2}$. Its steady state solution was calculated by Vinogradov and Willis for infinite parallel screw dislocations.\cite{vinogradov1,vinogradov2} Their motion was constrained to simple glide in one direction, so the difference between the system studied by them and the one considered in this paper is only the form of the generated stress field of the particles $\tau_\text{ind}$. They deduced an exact solution for the correlation function when there were only positive screw dislocations. However, in the case when both positive and negative screw dislocations were present, from the equation for the correlation functions they deduced a mathematical contradiction. Namely, the correlation functions had to tend to a constant nonzero value for large distances, which is obviously incorrect for real dislocation arrangements. They also showed that this contradiction was independent from the actual form of $\tau_\text{ind}$, which means that for edge dislocations the equations are not solvable either. We thus speculate that according to this result the cluster approximation is unphysical for the system of parallel edge dislocations.

Another traditional method for closing the hierarchy at this order is the Kirkwood superposition approximation\cite{kirkwood}, which was successfully used for instance in fluid mechanics (for a brief overview see, e.g., Ref.\ \onlinecite{balescu}). In the context of dislocations, it was first introduced by Zaiser \emph{et al}.\cite{zaiser1} but after investigating the resulting equations only a few basic conclusions were drawn.

The Kirkwood superposition approximation expresses the three-particle density functions in terms of the pair densities:
\begin{eqnarray}
&& \rho_3^{s_1, s_2, s_3}(\bm r_1, \bm r_2, \bm r_3, t) \nonumber \\
&& \quad \approx \frac{\rho_2^{s_1, s_2}(\bm r_1, \bm r_2, t) \rho_2^{s_2, s_3}(\bm r_2, \bm r_3, t) \rho_2^{s_3, s_1}(\bm r_3, \bm r_1, t)}{\rho_1^{s_1}(\bm r_1, t) \rho_1^{s_2}(\bm r_2, t) \rho_1^{s_3}(\bm r_3, t)} \nonumber \vspace*{50pt}\\
&& \quad = \rho_1^{s_1}(\bm r_1, t) \rho_1^{s_2}(\bm r_2, t) \rho_1^{s_3}(\bm r_3, t) \nonumber \\
&& \quad \quad \times [ 1 + d^{s_1,s_2}(\bm r_1, \bm r_2, t) ] [ 1 + d^{s_2,s_3}(\bm r_2, \bm r_3, t) ] \nonumber \\
&& \quad \quad \times [ 1 + d^{s_3,s_1}(\bm r_3, \bm r_1, t) ]. \label{eqn:kirkwood}
\end{eqnarray}
If we keep only the terms linear in the correlation functions $d^{s_1,s_2}$ this expression gives the same result as the ``cluster approximation'' (\ref{eqn:cluster}), i.e., their asymptotic behavior for large distances is identical. But as we will see, the nonlinear terms play an important role in the forthcoming derivation. We would like to emphasize that the Kirkwood factorization is indeed an approximation. For example, it does not fulfill the simple normalization condition, that if the number of particles is finite (let's say $N$), then $\int_{\mathbb{R}^2} \rho_3(.,.,\bm r, t) d^2 r = (N-2)\rho_2$. Thus it is expected that the superposition approximation can't give precise results for small distances.

\setlength{\arraycolsep}{.13889em}
In the following, we continue with substituting the approximation (\ref{eqn:kirkwood}) into the second order evolution equation (\ref{eqn:two_part_evol}). According to the previous results for the studied homogeneous system [Eqs.\ (\ref{eqn:const_rho}) and (\ref{eqn:rho_2_rel_coords})], the second term of (\ref{eqn:two_part_evol}) simplifies to
\begin{eqnarray}
&& \left( \frac{\partial}{\partial \bm r_1} - \frac{\partial}{\partial \bm r_2} \right) \left[ \rho_2^{s_1, s_2}(\bm r_1, \bm r_2, t) \bm F^{s_1, s_2}(\bm r_1 - \bm r_2) \right] \nonumber \\
&& \quad\quad = 2 s_1 s_2 \frac{\partial}{\partial \bm r_1} \left[ \bm b \rho_2^{s_1, s_2}(\bm r_1 - \bm r_2, t) \tau_\text{ind}(\bm r_1 - \bm r_2) \right].\quad\quad
\end{eqnarray}
For the third and the fourth terms of (\ref{eqn:two_part_evol}) one gets
\begin{widetext}
\begin{eqnarray}
&& \frac{\partial}{\partial \bm r_1} \sum_{s_3=\pm 1} \int\limits_{\mathbb R^2} \rho_3^{s_1, s_2, s_3}(\bm r_1, \bm r_2, \bm r_3, t) \bm F^{s_1, s_3}(\bm r_1 - \bm r_3) d^2 r_3 \nonumber \\
&& \quad = \frac{s_1}{\rho^{s_1} \rho^{s_2}} \frac{\partial}{\partial \bm r_1} \left[ \bm b \rho_2^{s_1, s_2}(\bm r_1 - \bm r_2, t) \sum_{s_3=\pm 1} \frac{s_3}{\rho^{s_3}} \int\limits_{\mathbb R^2} \rho_2^{s_2, s_3}(- \bm r_3, t) \rho_2^{s_3, s_1}(\bm r_3 - \bm r_1 + \bm r_2, t) \tau_{\text{ind}}(\bm r_1 - \bm r_2 - \bm r_3) d^2 r_3 \right], \quad
\end{eqnarray}
and
\begin{eqnarray}
&& \frac{\partial}{\partial \bm r_2} \sum_{s_3=\pm 1} \int\limits_{\mathbb R^2} \rho_3^{s_1, s_2, s_3}(\bm r_1, \bm r_2, \bm r_3, t) \bm F^{s_2, s_3}(\bm r_2 - \bm r_3) d^2 r_3 \nonumber \\
&& \quad = - \frac{s_2}{\rho^{s_1} \rho^{s_2}} \frac{\partial}{\partial \bm r_2} \left[ \bm b \rho_2^{s_1, s_2}(\bm r_1 - \bm r_2, t) \sum_{s_3=\pm 1} \frac{s_3}{\rho^{s_3}} \int\limits_{\mathbb R^2} \rho_2^{s_2, s_3}(\bm r_3 - \bm r_1 + \bm r_2, t) \rho_2^{s_3, s_1}(-\bm r_3, t) \tau_{\text{ind}}(\bm r_1 - \bm r_2 - \bm r_3) d^2 r_3 \right], \quad
\end{eqnarray}
where we also applied the $\tau_\text{ind}(-\bm r) = - \tau_\text{ind}(\bm r)$ relation. Finally, after introducing $\bm r := \bm r_1 - \bm r_2$, for the evolution equations of the second order one arrives at
\begin{eqnarray}
&& \frac{\partial \rho_2^{s_1,s_2}(\bm r, t)}{\partial t} + \frac{\partial}{\partial \bm r} \Bigg\{ \bm b \rho_2^{s_1, s_2}(\bm r, t) \Bigg[ 2 s_1 s_2 \tau_\text{ind}(\bm r) + (s_1 - s_2) \tau_\text{ext} + \frac{1}{\rho^{s_1} \rho^{s_2}} \sum_{s_3=\pm 1} \frac{s_3}{\rho^{s_3}} \int\limits_{\mathbb R^2} \Big( s_1 \rho_2^{s_3, s_2}(\bm r_3, t) \rho_2^{s_1, s_3}(\bm r - \bm r_3, t) \nonumber \\
&& \quad + s_2 \rho_2^{s_1, s_3}(\bm r_3, t) \rho_2^{s_3, s_2}(\bm r - \bm r_3, t) \Big) \tau_{\text{ind}}(\bm r - \bm r_3) d^2 r_3 \Bigg] \Bigg\} = 0.
\label{eqn:rho_2_hom}
\end{eqnarray}
\end{widetext}

The correlation functions, defined by Eq.\ (\ref{eqn:corr}), due to (\ref{eqn:const_rho}) and (\ref{eqn:rho_2_rel_coords}) simplify to
\begin{equation}
d^{s_1,s_2}(\bm r, t) = \frac{\rho_2^{s_1,s_2}(\bm r, t)}{\rho^{s_1} \rho^{s_2}} - 1.
\label{eqn:corr_hom}
\end{equation}
By substituting (\ref{eqn:corr_hom}) into (\ref{eqn:rho_2_hom}) one gets a closed set of equations for the two-particle correlation functions $d^{s_1,s_2}$.

In the rest of this paper it is assumed, that:
\begin{itemize}
\item The number of positive and negative signed dislocations are equal ($\rho^+ = \rho^-$).
\item The external shear stress is zero ($\tau_\text{ext} = 0$).
\end{itemize}
As it is explained in details in the Appendix, it follows that $d^{++}(\bm r, t) = d^{--}(\bm r, t)$ and $d^{+-}(\bm r, t) = d^{-+}(\bm r, t)$ for every $\bm r$ and $t$, meaning there are only two independent correlation functions: $d^{++}$ and $d^{+-}$. Under these conditions the evolution equations simplify to (for details see the Appendix):
\begin{equation}
\partial_t d^{++} + 2 \nabla \left[ \bm b (1+d^{++}) (\tau_\text{ind} + \tau_\text{sc}^h + \tau_b^h + \tau_a^h) \right] = 0, \label{eqn:evol_++}
\end{equation}
\begin{equation}
\partial_t d^{+-} + 2 \nabla \left[ \bm b (1+d^{+-}) (-\tau_\text{ind} - \tau_\text{sc}^h - \tau_b^h + \tau_a^h) \right] = 0, \label{eqn:evol_+-}
\end{equation}
where we have introduced the following terms having stress dimension:
\begin{eqnarray}
\tau_\text{sc}^h(\bm r, t) &:=& \rho^+ \int\limits_{\mathbb R^2} 2 d_d(\bm r', t) \tau_{\text{ind}}(\bm r - \bm r') d^2 r', \\
\tau_b^h(\bm r, t) &:=& \rho^+ \int\limits_{\mathbb R^2} 2 d_d(\bm r', t) d_s(\bm r - \bm r', t) \tau_{\text{ind}}(\bm r - \bm r') d^2 r', \nonumber \\
\\
\tau_a^h(\bm r, t) &:=& \rho^+ \int\limits_{\mathbb R^2} 2 d_s(\bm r', t) d_d(\bm r - \bm r', t) \tau_{\text{ind}}(\bm r - \bm r') d^2 r' \nonumber \\
\end{eqnarray}
with $d_s := (d^{++} + d^{+-})/2$ and $d_d := (d^{++} - d^{+-})/2$.

Dislocations of opposite signs should move at each given point with equal velocities but in opposite directions, which implies
\begin{equation}
\tau_a^h(\bm r, t) = 0 \label{eqn:tau_a_zero}
\end{equation}
for every $\bm r$ and $t$. This argument is identical to that in the appendix of Ref.\ \onlinecite{groma2}, where a similar stress term is omitted.

The final evolution equations (\ref{eqn:evol_++}) and (\ref{eqn:evol_+-}) with (\ref{eqn:tau_a_zero}) are very similar to those obtained by Groma \emph{et al.}\ for the one particle dislocation densities.\cite{groma2} To emphasize the analogy even more, let us introduce the following notations:
\begin{eqnarray}
\rho^h(\bm r, t) &:=& \rho^+ \left[ 2 + d^{++}(\bm r, t) + d^{+-}(\bm r, t) \right] \nonumber \\
&=& 2 \rho^+ \left[ 1 + d_s(\bm r, t) \right], \label{eqn:rho_h_evol} \\
\kappa^h(\bm r, t) &:=& \rho^+ \left[ d^{++}(\bm r, t) - d^{+-}(\bm r, t) \right] \nonumber \\
&=& 2 \rho^+ d_d(\bm r, t). \label{eqn:kappa_h_evol}
\end{eqnarray}
(Here and previously the `h' superscript refers to the homogeneous system.) Although these quantities have density dimensions, they are auxiliary quantities and do not carry the meaning of single particle densities. The evolution equations for these newly introduced quantities can be written as
\begin{eqnarray}
\partial_t \rho^h + 2 \nabla \left[ \bm b \kappa^h (\tau_\text{ind} + \tau_\text{sc}^h + \tau_b^h) \right] &=& 0, \label{eqn:evol_rho_h} \\
\partial_t \kappa^h + 2 \nabla \left[ \bm b \rho^h (\tau_\text{ind} + \tau_\text{sc}^h + \tau_b^h) \right] &=& 0 \label{eqn:evol_kappa_h} \qquad
\end{eqnarray}
with
\begin{eqnarray}
\tau_\text{sc}^h(\bm r, t) &=& \int\limits_{\mathbb R^2} \kappa^h(\bm r', t) \tau_{\text{ind}}(\bm r - \bm r') d^2 r', \\
\tau_b^h(\bm r, t) &=& \int\limits_{\mathbb R^2} \kappa^h(\bm r', t) d_s(\bm r - \bm r', t) \tau_{\text{ind}}(\bm r - \bm r') d^2 r'. \quad\quad
\end{eqnarray}

These are the equations which govern the evolution of dislocation-dislocation correlation functions. Instead of their numerical solution, we will prove their correctness by an analogy with the phenomena of screening of an individual dislocation, which is discussed in the following section.

\section{Dislocation density fields of a screened external dislocation}
\label{sec:debye}

The stress field of a single dislocation decays as $1/r$. This implies that if the distribution of the dislocations was completely random in a crystal, then the elastic energy per unit volume would diverge logarithmically with the crystal size.\cite{wilkens} Since this cannot be observed in real systems, the only possible solution is, that dislocations arrange themselves in a correlated way which screens out their long range effect. The phenomenon lends itself to the analogy with Debye screening in Coulomb systems.

In order to address the problem, Groma {\em et al.}\cite{groma3, groma5} studied the induced geometrically necessary dislocation density around a single external edge dislocation in 2D, proposed an equation for the stress potential in equilibrium and gave analytic solution for the infinite plane, provided the total density was constant and much larger than the geometrically necessary dislocation density. While this is not the original question of the correlation function, within linear response theory, however, the result is expected to be valid also to the correlation problem. It was indeed shown that the screened dislocation's stress field decays faster than that of the unscreened one's, in the direction perpendicular to the Burgers vector by a power law, and in all other directions exponentially. This result was actually compared with correlation simulations, mostly because this was numerically simpler than the computation of the induced field by an external dislocation, and in the direction of the power decay not only the exponent but the entire shape predicted by theory was recovered in the simulation. However, the exponential decay along the $x$ axis was not seen as in the theoretical solution, mostly because of the special conditions of dislocation motion on a torus as realized in the simulation. In any event, the equivalence of response to an external dislocation and the correlation in the absence of a fixed, external one, was taken as granted, precisely the problem addressed in the present paper.

An important peculiar aspect of screening of dislocations has to be emphasized. First, we recall that in a thermal system like Coulomb plasma linear response theory provides enough ground to expect the similarity between screening of an external charge and that in the correlation. On the other hand, the screening problem of dislocations already emerges at zero physical temperature. Now the role of the temperature in keeping oppositely charged particles at distance is taken over by the constraint to slip axes, so here an effective temperature parameter arises that can be determined from comparison with simulation.\cite{groma3, groma5} Now given the fact that we only have a temperature parameter (which directly appears in the effective thermodynamic potential) but not usual thermodynamics, it is by far not obvious that the two types of screening, namely, the one of an external dislocation, and the one appearing in the correlation function, should maintain the same type of equivalence as if Boltzmannian thermodynamics were valid. Hence, in the case of dislocations it remains an open problem, what the relation is between screening by response to an external effect and screening in the correlations.

In the previous section we constructed the equations of motion for correlations, now we do the same for the one-particle densities in the presence of an external dislocation. As a result of the stress field generated by the inserted object, the positions of the other dislocations change and a new relaxed state evolves. In it, for instance, one particle dislocation densities won't be constant any more. In this section, the time evolution of these functions is investigated.

Due to the spatially varying extra force acting on the dislocations the system is not translation invariant any more.
So, in the case of screening of an external dislocation, contrary to the case in the previous section, the system is spatially inhomogeneous. The evolution of the one-particle densities is described by the first member of the BBGKY hierarchy (\ref{eqn:one_part_evol}):
\begin{eqnarray}
&&\frac{\partial \rho_1^{s_1}(\bm r_1, t)}{\partial t} \nonumber \\
&& \quad + \frac{\partial}{\partial \bm r_1} \sum_{s_2=\pm 1} \int\limits_{\mathbb R^2} \rho_2^{s_1,s_2}(\bm r_1, \bm r_2, t) \bm F^{s_1,s_2}(\bm r_1 - \bm r_2) d^2 r_2 \nonumber \\
&& \quad + \frac{\partial}{\partial \bm r_1} \left[ \rho_1^{s_1}(\bm r_1, t) \bm F_\text{scr}^{s_1}(\bm r_1) \right] = 0, \label{eqn:one_part_evol_screening}
\end{eqnarray}
where we introduced the notation $\bm F_\text{scr}^{s}(\bm r) := s b \bm b \tau_\text{ind}(\bm r)$ with $b := |\bm b|$ for the stress field of the extra dislocation.

According to the deduction in Ref.\ \onlinecite{groma2}, Eq.~(\ref{eqn:one_part_evol_screening}) can be cast into the following form
\begin{eqnarray}
\partial_t \rho + \nabla \left[ \bm b \kappa (\tau_\text{ind} + \tau_\text{sc} - \tau_f + \tau_b) \right] &=& 0, \label{eqn:evol_rho} \\
\partial_t \kappa + \nabla \left[ \bm b \rho (\tau_\text{ind} + \tau_\text{sc} - \tau_f + \tau_b) \right] &=& 0, \label{eqn:evol_kappa} \qquad
\end{eqnarray}
where we have introduced the
\begin{equation}
\rho(\bm r, t) := \rho^+(\bm r, t) + \rho^-(\bm r, t) \label{eqn:rho}
\end{equation}
and
\begin{equation}
\kappa(\bm r, t) := \rho^+(\bm r, t) - \rho^-(\bm r, t), \label{eqn:kappa}
\end{equation}
total and geometrically necessary dislocation densities, respectively, and
\begin{eqnarray}
\tau_\text{sc}(\bm r, t) &:=& \int\limits_{\mathbb R^2} \kappa(\bm r', t) \tau_{\text{ind}}(\bm r - \bm r') d^2 r', \\
\tau_b(\bm r, t) &:=& \int\limits_{\mathbb R^2} \kappa(\bm r', t) \tilde{d}(\bm r, \bm r', t) \tau_{\text{ind}}(\bm r - \bm r') d^2 r', \qquad \label{eqn:tau_b} \\
\tau_f(\bm r, t) &:=& \frac12 \int\limits_{\mathbb R^2} \rho(\bm r', t) \tilde{d}_a(\bm r, \bm r', t) \tau_{\text{ind}}(\bm r - \bm r') d^2 r', \nonumber \\
\label{eqn:tau_f}
\end{eqnarray}
with $\tilde{d} := (\tilde{d}^{++} + \tilde{d}^{--} + \tilde{d}^{+-} + \tilde{d}^{-+})/4$ and $\tilde{d}_a := (\tilde{d}^{+-} - \tilde{d}^{-+})/2$. We note that in this inhomogeneous case the correlation functions do not depend on the difference of their arguments, meaning one has to remain at the general definition (\ref{eqn:corr}). Here and in the rest of this paper $(\tilde{\cdot})$ indicates that the correlation function has two spatial variables. During the derivation of Eqs.~(\ref{eqn:evol_rho}, \ref{eqn:evol_kappa}) no approximations have been made.

If the correlation functions were short-range, it could be approximated that they depend only on the relative coordinate $\bm r_1 - \bm r_2$:
\begin{eqnarray}
&& \rho_2^{s_1,s_2}(\bm r_1, \bm r_2, t) \nonumber \\
&& \quad = \rho_1^{s_1}(\bm r_1, t) \rho_1^{s_2}(\bm r_2, t) \left[ 1 + d^{s_1,s_2}(\bm r_1 - \bm r_2, t) \right]. \quad
\label{eqn:short_corr}
\end{eqnarray}
Here the correlation function $d^{s_1,s_2}$ can be taken from homogeneous systems.\cite{groma2} The assumed shortness of dislocation-dislocation correlations was proved on discrete dislocation simulation results earlier.\cite{zaiser1} It was found that the correlation functions decay to zero exponentially within a few average dislocation spacings.

Because of the short-range nature of dislocation-dislocation correlations, the $\kappa$ and $\rho$ functions can be approximated by their Taylor expansions in the integrals of Eqs.\ (\ref{eqn:tau_b}) and (\ref{eqn:tau_f}). After keeping only the first non-vanishing terms Groma \emph{et al}.\ arrived at
\begin{equation}
\tau_b(\bm r, t) = -\frac{\mu}{2 \pi (1-\nu)} D_d \frac{b}{\rho(\bm r, t)} \partial_x \kappa(\bm r, t) \label{eqn:b_stress}
\end{equation}
 and
\begin{equation}
\tau_f(\bm r, t) = \frac{\mu}{4 \pi (1-\nu)} C_d b \sqrt{\rho(\bm r, t)}, \label{eqn:f_stress}
\end{equation}
where $D_d$ and $C_d$ are positive constants.\cite{groma2} The term $\tau_b$ is a gradient like contribution to the stress, and is called ``back-stress'', while $\tau_f$ can be interpreted as a local flow-stress. The physical correctness of these approximations was proved by discrete dislocation simulations.\cite{groma2, yefimov1, yefimov2}

The description of Debye screening of an external dislocations was based on a thermodynamic potential-like functional, from which the evolution equations (\ref{eqn:evol_rho}, \ref{eqn:evol_kappa}) for $\kappa$ and $\rho$ can be derived.\cite{groma3, groma5} In it $\tau_b$ is approximated as (\ref{eqn:b_stress}) and $\tau_f$ is omitted. In the case of $\rho = \text{const.}$ the analytical solution of the static case was given and compared with numerical simulations .\cite{groma3}

In what follows we shall see that, to establish the connection between the evolution of correlation functions and the screening field of a single external dislocations, we will not need the approximations (\ref{eqn:short_corr}), (\ref{eqn:b_stress}), and (\ref{eqn:f_stress}) for the correlation functions, $\tau_b$, and $\tau_f$, respectively, rather can keep the more general definitions (\ref{eqn:tau_b}) and (\ref{eqn:tau_f}).

\section{Connection between the screened dislocation densities and the correlation functions}
\label{sec:conn}

In the previous two sections we have derived evolution equations for correlation functions in a homogeneous system, within the Kirkwood approximation, as well as for induced dislocation densities around a fixed, external, dislocation. A main message of this paper is that the deduced equations for the two different cases, namely (\ref{eqn:rho_h_evol},\ref{eqn:kappa_h_evol}) and (\ref{eqn:evol_rho},\ref{eqn:evol_kappa}), are very similar. More is true, however, they become identical, if special equalities hold for various $d$-s and time is appropriately rescaled. In detail, these conditions are the following:

\begin{itemize}
\item If the numbers of $+$ and $-$ dislocations are the same, then in the absence of external stresses $d^{+-}(\bm r,t) = d^{-+}(\bm r,t)$ and $d^{++}(\bm r,t) = d^{--}(\bm r,t)$. So, if the general correlation functions are approximated by the ones taken from homogeneous systems, i.e.,
\begin{equation}
\tilde{d}^{s_1, s_2}(\bm r_1, \bm r_2, t) = d^{s_1, s_2}(\bm r_1 - \bm r_2, t), \label{eqn:lin_resp}
\end{equation}
then $\tilde{d}_a(\bm r_1, \bm r_2, t) = 0$ and $\tilde{d}(\bm r_1, \bm r_2, t) = d_s(\bm r_1 - \bm r_2, t)$. In this case the ``flow-stress'' term $\tau_f$ disappears from (\ref{eqn:evol_rho},\ref{eqn:evol_kappa}), while it was never present in Eqs.\ (\ref{eqn:rho_h_evol}, \ref{eqn:kappa_h_evol}). Moreover, the $\tau_b$ in Eqs.\ (\ref{eqn:evol_rho},\ref{eqn:evol_kappa}) becomes equal to the $\tau_b^h$ of (\ref{eqn:rho_h_evol},\ref{eqn:kappa_h_evol}).


\emph{The main theoretical result of this paper immediately follows. Namely, if one takes a solution of Eqs.\ (\ref{eqn:rho_h_evol},\ref{eqn:kappa_h_evol}) for $d^{++}(\bm r,t)$ and $d^{+-}(\bm r,t)$, next substitutes these into (\ref{eqn:evol_rho},\ref{eqn:evol_kappa}), then the latter equations will have a solution identical with the one of (\ref{eqn:rho_h_evol},\ref{eqn:kappa_h_evol}).}

Physically, the equality $\tilde{d}(\bm r_1, \bm r_2, t) = d_s(\bm r_1 - \bm r_2, t)$ corresponds to the approximation when the effect of the stress by the external dislocation in Eqs.\ (\ref{eqn:evol_rho},\ref{eqn:evol_kappa}) are neglected in the correlation functions. This condition amounts to taking the external effect as a first order perturbation, in perfect analogy of linear response theory valid for conventional thermal systems.

\item The divergence-like terms of (\ref{eqn:rho_h_evol}) and (\ref{eqn:kappa_h_evol}) are multiplied by two, which only affects the time scale of the process. The appearance of this factor can be well understood if we consider two dislocations moving in each other's stress fields. They relax into the same configuration either we keep the position of the first dislocation fixed or not. The only difference is, that the relaxation in the first case (which corresponds to screening of an individual dislocation) lasts twice as long as in the case when both dislocations can glide freely (case of the evolution of correlation functions).
\end{itemize}

To sum up, since at $t=0$ $\rho(\bm r,0) = \rho^h(\bm r,0) = \rho^+ + \rho^-$ and $\kappa(\bm r,0) = \kappa^h(\bm r,0) = 0$, it follows that
\begin{equation}
\rho(\bm r, 2t) = \rho^h(\bm r, t) = \rho^+ \left[ 2 + d^{++}(\bm r, t) + d^{+-}(\bm r, t) \right] \label{eqn:rho_rho_h}
\end{equation}
and
\begin{equation}
\kappa(\bm r, 2t) = \kappa^h(\bm r, t) = \rho^+ \left[ d^{++}(\bm r, t) - d^{+-}(\bm r, t) \right] \label{eqn:kappa_kappa_h}
\end{equation}
for every $\bm r$ and $t$. We would like to stress, that during the derivation of this statement we made approximations for both $(\rho^h, \kappa^h)$ and for $(\rho, \kappa)$ functions. For the first set the Kirkwood superposition approximation [Eq.\ (\ref{eqn:kirkwood})], and for the second linear response was assumed, meaning general correlation functions were approximated by ones taken from infinite homogeneous systems [Eq.~(\ref{eqn:lin_resp})]. If the comparison of the numerically obtained functions confirms the validity of (\ref{eqn:rho_rho_h}) and (\ref{eqn:kappa_kappa_h}), then it would also confirm the applicability of the Kirkwood superposition approximation. In the following section this comparison is discussed.

\section{Simulation results}
\label{sec:num_res}

\begin{figure}[!ht]
\begin{center}
\includegraphics[height=7.5cm, angle=0]{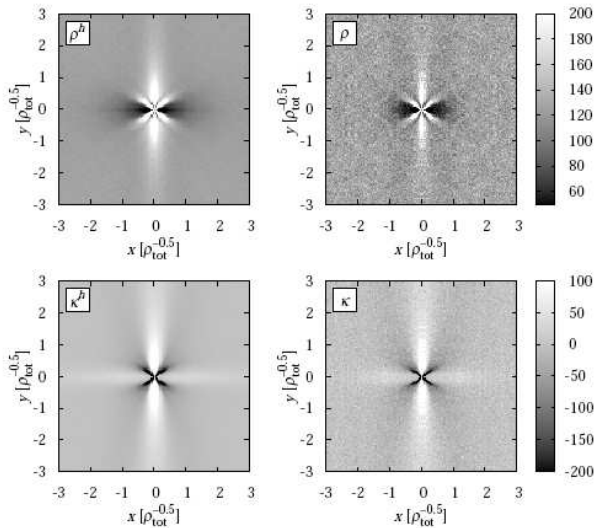}
\end{center}
\caption{The $\rho^h$, $\rho$, $\kappa^h$, and $\kappa$ functions in the relaxed state obtained from numerical simulations. The distances are measured in $\rho_\text{tot}^{-0.5}$ average dislocation spacings, where $\rho_\text{tot} := \rho^+ + \rho^-$ is the total dislocation density. \label{fig:rho_kappa_relaxed}}
\end{figure}

To measure $\rho^h$, $\rho$, $\kappa^h$, and $\kappa$ numerically, we first performed more than 13000 different relaxations. In each of them at the beginning 128 dislocations (64 with positive and 64 with negative sign) were distributed randomly in an $L \times L$ square-like domain with periodic boundary conditions, and then they relaxed to a steady state. The $d^{++}$ and $d^{+-}$ correlation functions and then the $\rho^h$ and $\kappa^h$ quantities were determined from these configurations by counting the relative coordinates of the dislocations. They are plotted in the first column of Fig.\ \ref{fig:rho_kappa_relaxed}.

Afterwards, a new dislocation with positive sign was placed at the center of the simulation area in each configuration, and the systems were let to relax again while the extra dislocation was fixed. These simulations correspond to the phenomena of Debye screening of an external dislocation. The $\rho^+$ and $\rho^-$ densities were obtained by counting the number of positive and negative dislocations falling into each cell of a $256\times256$ mesh. Then $\rho$ and $\kappa$ functions defined by (\ref{eqn:rho}) and (\ref{eqn:kappa}) were calculated. They are also plotted in Fig.\ \ref{fig:rho_kappa_relaxed}.

\begin{figure}[!ht]
\begin{center}
\hspace*{-1cm}
\includegraphics[width=4cm, angle=-90]{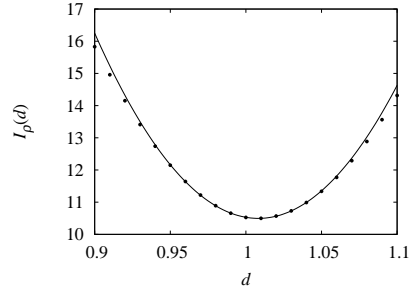}
\end{center}
\caption{The $I_\rho$ function calculated numerically at different values (data points). A second order polynomial was fitted around the minimum of the points (solid line). \label{fig:i_rho}}
\end{figure}

The two functions, according to our expectations, are quite similar. The difference is that in the case of $\kappa$ and $\rho$ the amplitude of the noise is much higher. The simple reason for this is that a relaxed system of $N$ dislocations gives $N(N-1)$ data (the relative coordinates) during the calculation of the correlation functions but only $N$ data (the positions) when calculating the densities. In order to make a comparison between these functions the following function was considered:
\begin{equation}
	I_\rho(d) := \int\limits_A \left| \rho(\bm r) - d\, \rho^h(\bm r) \right| d^2 r. \label{eqn:integral}
\end{equation}
The integral of a noise is nearly constant so if $\rho$ and $\rho^h$ are equal except for the noise then this quantity must have a minimum at $d=1$. As it was mentioned in Sec.\ \ref{sec:time_evol}, the Kirkwood superposition approximation is not expected to be correct for distances smaller than an average dislocation spacing, and so, Eqs.~(\ref{eqn:rho_rho_h}) and (\ref{eqn:kappa_kappa_h}) should apply only for $r = |\bm r| \gtrsim \rho_\text{tot}^{-0.5}$ ($\rho_\text{tot} := \rho^+ + \rho^-$ and $\rho_\text{tot}^{-0.5}$ is the average dislocation spacing). Accordingly, it is meaningful to define the domain of the integration $A$ so that a circle of radius $0.5 \rho_\text{tot}^{-0.5}$ around the origin is left out from it. The numerically obtained function $I_\rho$ is plotted in Fig.\ \ref{fig:i_rho}. As it can be seen, $I_\rho$ indeed has a minimum around $d=1$. This confirms that the $\rho^h$ and $\rho$ functions can be considered equal in first approximation. The repetition of this calculation for $\kappa^h$ and $\kappa$ leads to similar results.

\begin{figure*}[!ht]
\begin{center}
\includegraphics[height=7.5cm, angle=0]{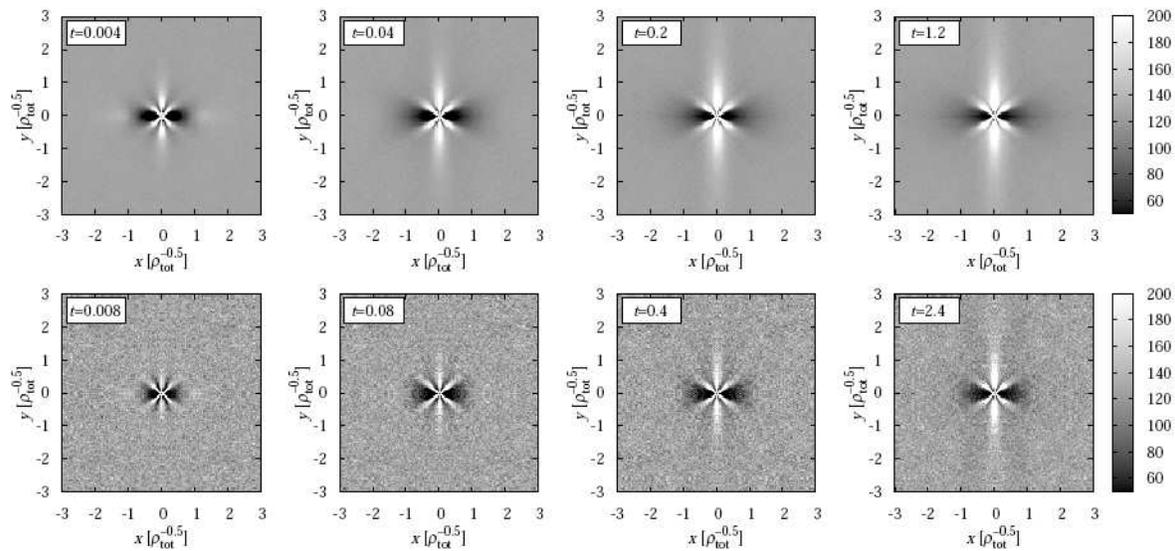}
\end{center}
\caption{The evolution of $\rho^h$ (first row) and $\rho$ (second row). The time $t$ is measured in $\frac{b^2 \mu}{2 \pi (1 - \nu) B L^2}$ dimensionless units, where $B$ is the dislocation mobility. According to Eq.~(\ref{eqn:rho_rho_h}) $\rho^h(\bm r,t)$ and $\rho(\bm r,2t)$ are compared. \label{fig:evol_rho}}
\end{figure*}

\begin{figure*}[!ht]
\begin{center}
\includegraphics[height=7.5cm, angle=0]{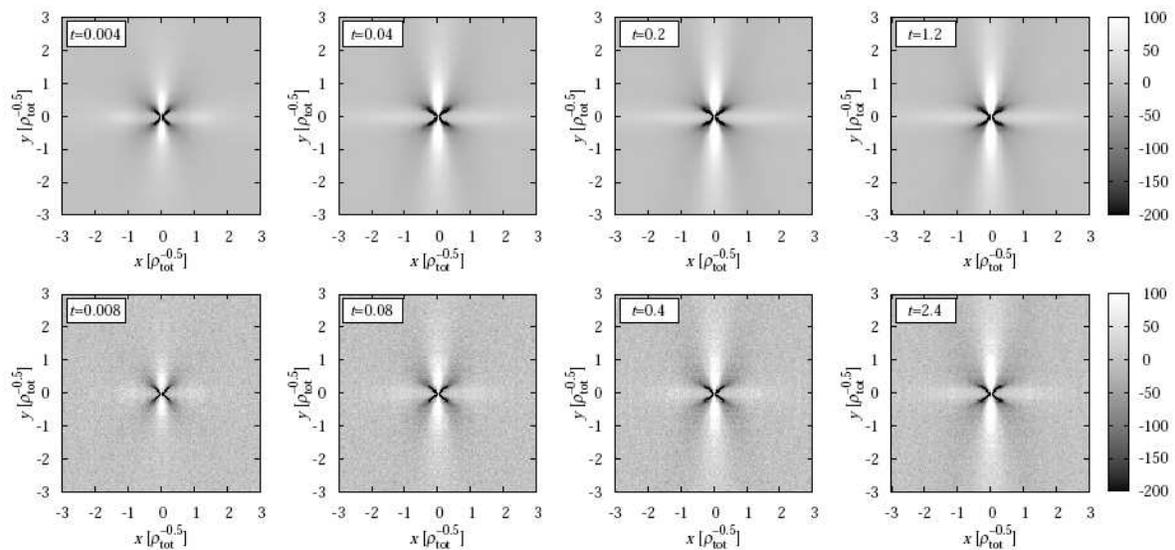}
\end{center}
\caption{The evolution of $\kappa^h$ (first row) and $\kappa$ (second row). The time $t$ is measured in $\frac{b^2 \mu}{2 \pi (1 - \nu) B L^2}$ dimensionless units, where $B$ is the dislocation mobility. According to Eq.~(\ref{eqn:kappa_kappa_h}) $\kappa^h(\bm r,t)$ and $\kappa(\bm r,2t)$ are compared. \label{fig:evol_kappa}}
\end{figure*}

In Fig.\ \ref{fig:evol_rho} $\rho^h$ and $\rho$ functions can be seen at different times obtained from the numerical simulations. According to (\ref{eqn:rho_rho_h}), $\rho^h(\bm r, t)$ was compared with $\rho(\bm r,2t)$, because of the factor of $2$ appearing in the evolution equations of the correlation functions in Eqs.~(\ref{eqn:evol_rho_h}) and (\ref{eqn:evol_kappa_h}). The same can be seen for the $\kappa$ functions in Fig.\ \ref{fig:evol_kappa}. To summarize, it can be stated that (\ref{eqn:rho_rho_h}) and (\ref{eqn:kappa_kappa_h}) are, at least in first order, fulfilled.

\section{Conclusions}

In this paper we derived evolution equations for the correlation functions of long parallel edge dislocations. We started our analysis at the BBGKY hierarchy of dislocation many-body densities, which was deduced earlier from the equations of motion of individual dislocations.\cite{groma1} As a closure approximation for the hierarchy, the Kirkwood superposition approximation was used. Our investigations showed that the evolution of the correlation functions and the dislocation density fields evolving around a screened dislocation are closely related (according to Eqs.\ (\ref{eqn:rho_rho_h}) and (\ref{eqn:kappa_kappa_h}) only the time scale of the processes is different). In Sec.\ \ref{sec:num_res} this statement was confirmed by numerical simulation.

The Kirkwood superposition approximation (\ref{eqn:kirkwood}) gives the three-particle dislocation density in terms of the pair correlations. Contrary to the cluster approximation (\ref{eqn:cluster}), it contains quadratic and cubic terms of the correlation function beside the linear one. As it was mentioned in Sec.\ \ref{sec:time_evol} these higher-order terms play an important role. According to our analysis presented in this paper, the omission of the cubic term leads to the disappearance of the back-stress term $\tau_b^h$ in the evolution equations (\ref{eqn:evol_rho_h}) and (\ref{eqn:evol_kappa_h}). This corresponds to the mean-field approximation where dislocation correlations are neglected resulting that the stress acting on dislocations is simply the self-consistent field $\tau_\text{sc}^h$.\cite{groma1, groma4}

If one neglects the cubic and quadratic terms (which is the cluster approximation), not only the back-stress $\tau_b^h$, but also the self-consistent field $\tau_\text{sc}^h$ disappears, which is obviously unphysical. We mentioned in Sec.\ \ref{sec:time_evol} that the cluster approximation (\ref{eqn:cluster}) used by Vinogradov and Willis leads to a mathematical contradiction in the case of edge dislocations of opposite signs.\cite{vinogradov1,vinogradov2} This result is in complete agreement with our analysis.

The Kirkwood superposition approximation leads to the evolution equations (\ref{eqn:evol_rho_h}) and (\ref{eqn:evol_kappa_h}). Their correctness was proved implicitly by the facts that (i) analogy was found with the one-particle density evolution equations and (ii) numerical work confirmed the similitude between the evolution of the one- and two-particle distributions. Another way of testing the deduced integro-differential evolution equations could be their numerical solution. This is out of the scope of the present paper, and requires further numerical work.

\section*{Acknowledgment}

The financial support of the Hungarian Scientific Research Fund (OTKA) under Contract No.\ K 67778 and of the European Community's Human Potential Programme under Contract No.\ NMP3-CT-2006-017105 [DIGIMAT] are gratefully
acknowledged.

\appendix*
\section{}

In this section we derive evolution equations for the correlation functions [which were defined by Eq.\ (\ref{eqn:corr})] starting from the evolution equation (\ref{eqn:rho_2_hom}) of $\rho_2^{s_1,s_2}$. First, we assume that the densities of positive and negative dislocations are equal ($\rho^+ = \rho^-$). It can be noted that in this case if $\rho_2^{++}(\bm r,t) = \rho_2^{--}(\bm r,t)$ for any $t$, then the second term of (\ref{eqn:rho_2_hom}) is equal for both $\rho_2^{++}$ and $\rho_2^{--}$. This means, that the evolutions of $\rho_2^{++}$ and $\rho_2^{--}$ are identical, hence $\rho_2^{++}(\bm r,t) = \rho_2^{--}(\bm r,t)$ and therefore $d^{++}(\bm r,t) = d^{--}(\bm r,t)$ for every $t$. For this reason one only has to deal with $d^{++}$, $d^{+-}$, and $d^{-+}$.

Let us introduce
\begin{equation}
d_p := (d^{+-} + d^{-+})/2
\end{equation}
and
\begin{equation}
d_a := (d^{+-} - d^{-+})/2,
\end{equation}
namely the symmetric and the antisymmetric part of $d^{+-}$. Furthermore, we define $d_s$ and $d_d$ as the half of the sum and the difference of $d^{++}$ and $d_p$, respectively:
\begin{equation}
d_s := (d^{++} + d_p)/2
\end{equation}
and
\begin{equation}
d_d := (d^{++} - d_p)/2.
\end{equation}
It is also useful to introduce the following quantities having stress dimensions:
\setlength{\arraycolsep}{0em}
\begin{eqnarray}
&& \tau_\text{sc}^h(\bm r, t) := \rho^+ \int\limits_{\mathbb R^2} 2 d_d(\bm r', t) \tau_{\text{ind}}(\bm r - \bm r') d^2 r', \\
&& \tau_b^h(\bm r, t) := \rho^+ \int\limits_{\mathbb R^2} 2 d_d(\bm r', t) d_s(\bm r - \bm r', t) \tau_{\text{ind}}(\bm r - \bm r') d^2 r', \nonumber \\
\\
&& \tau_f^h(\bm r, t) := -\frac12 \rho^+ \int\limits_{\mathbb R^2} d_a(\bm r', t) d_a(\bm r - \bm r', t) \nonumber \\
&& \qquad \times \tau_{\text{ind}}(\bm r - \bm r') d^2 r', \\
&& \tau_a^h(\bm r, t) := \rho^+ \int\limits_{\mathbb R^2} \bigg[ 2 d_s(\bm r', t) d_d(\bm r - \bm r', t) \nonumber \\
&& \qquad + \frac12 d_a(\bm r', t) d_a(\bm r - \bm r', t) \bigg] \tau_{\text{ind}}(\bm r - \bm r') d^2 r', \\
&& \tau_p^h(\bm r, t) := \rho^+ \int\limits_{\mathbb R^2} \Big\{ d_a(\bm r', t) \left[ 1 + d^{++}(\bm r - \bm r', t) \right] \nonumber \\
&& \qquad - \left[ 1 + d^{++}(\bm r', t) \right] d_a(\bm r - \bm r', t) \Big\} \tau_{\text{ind}}(\bm r - \bm r') d^2 r'. \nonumber \\
\end{eqnarray}
With these new functions, from the evolution equations (\ref{eqn:rho_2_hom}) for the correlation functions one obtains:
\begin{eqnarray}
&& \partial_t d^{++} + 2 \nabla \big[ \bm b (1+d^{++}) \nonumber \\
&& \quad \times (\tau_\text{ind} + \tau_\text{sc}^h + \tau_b^h - \tau_f^h + \tau_a^h) \big] = 0, \label{eqn:evol_gen_dpp} \\
&& \partial_t d^{+-} + 2 \nabla \big[ \bm b (1+d^{+-}) \nonumber \\
&& \quad \times (-\tau_\text{ind} + \tau_\text{ext} - \tau_\text{sc}^h - \tau_b^h + \tau_f^h + \tau_a^h + \tau_p^h) \big] = 0, \nonumber \\
\\
&& \partial_t d^{-+} + 2 \nabla \big[ \bm b (1+d^{-+}) \nonumber \\
&& \quad \times (-\tau_\text{ind} - \tau_\text{ext} - \tau_\text{sc}^h - \tau_b^h + \tau_f^h + \tau_a^h - \tau_p^h) \big] = 0. \nonumber \\
\label{eqn:evol_gen_dmp}
\end{eqnarray}
This result can be checked by simple substitution.

In the absence of external stress ($\tau_\text{ext} = 0$) the general equations (\ref{eqn:evol_gen_dpp} - \ref{eqn:evol_gen_dmp}) can be simplified because for symmetry reasons $d^{+-}(\bm r, t) = d^{-+}(\bm r,t)$ for every $t$. In other words, at zero external stress if one alters the sign of all dislocations it does not affect the behavior of the system. This means that $d_a = 0$ and so $\tau_f^h = \tau_p^h = 0$. In this case the resulting evolution equations are:
\begin{eqnarray}
&& \partial_t d^{++} + 2 \nabla \left[ \bm b (1+d^{++}) (\tau_\text{ind} + \tau_\text{sc}^h + \tau_b^h + \tau_a^h) \right] = 0, \nonumber \\
\\
&& \partial_t d^{+-} + 2 \nabla \left[ \bm b (1+d^{+-}) (-\tau_\text{ind} - \tau_\text{sc}^h - \tau_b^h + \tau_a^h) \right] = 0. \nonumber \\
\end{eqnarray}

\end{document}